\documentclass[reprint,amsmath,amssymb,twocolumn,aps,prl]{revtex4-2}
\usepackage{graphicx}
\usepackage{dcolumn}
\usepackage{bm}
\usepackage{glossaries}
\usepackage{xcolor}

\newglossaryentry{invTemp}%
{%
  name={\ensuremath{\beta}},
  description={The inverse temperature, multiplied with the Boltzmann factor: $\beta = \frac{1}{k_B T}$},
  sort={Temp}
}

\newglossaryentry{sysVol}%
{%
  name={\ensuremath{\Omega}},
  description={The system volume. That is the volume of the crystal, usually},
  sort={Vol}
}

\newglossaryentry{ucVol}%
{%
  name={\ensuremath{\Omega_c}},
  description={The unit cell volume. That is the volume of the crystal or the supercell},
  sort={Vol}
}

\newglossaryentry{bzVol}%
{%
  name={\ensuremath{\Omega_\text{BZ}}},
  description={The Brillouin zone volume},
  sort={Vol}
}

\newglossaryentry{IFC}%
{%
  name={\ensuremath{\Phi}},
  description={The interatomic force constant},
  sort={Vol}
}

\newglossaryentry{IFCq}%
{%
  name={\ensuremath{\tilde{\Phi}}},
  description={The interatomic force constant, in inverse space},
  sort={Vol}
}

\newglossaryentry{epiC}%
{%
  name={\ensuremath{g}},
  description={An electron-phonon coupling element},
  sort={Vol}
}

\newglossaryentry{uci}%
{%
  name={\ensuremath{p}},
  description={The unit cell index},
  sort={Vol}
}

\newglossaryentry{nid}%
{%
  name={\ensuremath{\kappa}},
  description={Nucleus index in unit cell},
  sort={Vol}
}

\newglossaryentry{dir}%
{%
  name={\ensuremath{\alpha}},
  description={Placeholder for a summed direction},
  sort={Vol}
}

\newglossaryentry{part1}%
{%
  name={\ensuremath{l}},
  description={First particle index},
  sort={Vol}
}

\newglossaryentry{part2}%
{%
  name={\ensuremath{k}},
  description={Second particle index},
  sort={Vol}
}

\newglossaryentry{moi}%
{%
  name={\ensuremath{\nu}},
  description={Phonon mode index},
  sort={Vol}
}

\newglossaryentry{ucv}%
{
  name={\ensuremath{\mathbf{R}}},
  description={Lattice vector},
  sort={Vol}
}

\newglossaryentry{eqPos}%
{%
  name={\ensuremath{\mathbf{x}}},
  description={The equilibrium position of a particle in the crystal},
  sort={Loc}
}

\newglossaryentry{eqPosd}%
{%
  name={\ensuremath{x}},
  description={A component of the equilibrium position of a particle in the crystal},
  sort={Loc}
}

\newglossaryentry{specF}%
{%
  name={\ensuremath{\mathcal{S}}},
  description={Spectral function of an operator of Green's function},
  sort={Loc}
}

\newglossaryentry{occB}%
{%
  name={\ensuremath{n_B}},
  description={Boson occupation number},
  sort={Loc}
}

\newglossaryentry{occF}%
{%
  name={\ensuremath{n_F}},
  description={Fermion occupation number},
  sort={Loc}
}

\newglossaryentry{inf}%
{%
  name={\ensuremath{0^+}},
  description={Positive infinitesimal},
  sort={Loc}
}

\newglossaryentry{im}%
{%
  name={\ensuremath{\text{Im}}},
  description={Imaginary part of a number},
  sort={Loc}
}

\newglossaryentry{re}%
{%
  name={\ensuremath{\text{Re}}},
  description={Real part of a number},
  sort={Loc}
}

\newglossaryentry{selfE}%
{%
  name={\ensuremath{\Sigma}},
  description={Self-energy of a phonon},
  sort={Loc}
}

\newglossaryentry{eigene}%
{%
  name={\ensuremath{\mathcal{E}}},
  description={Eigenenergy of a state, usually with greek subletter},
  sort={Loc}
}

\newglossaryentry{prin}%
{%
  name={\ensuremath{\mathcal{P}}},
  description={Principal value of an integral. If standalone, the integration is implied},
  sort={Loc}
}

\newglossaryentry{res}%
{%
  name={\ensuremath{\text{Res}}},
  description={Residue of a function at some point},
  sort={Loc}
}

\newglossaryentry{tr}%
{%
  name={\ensuremath{\text{Tr}}},
  description={The trace of an operator / matrix},
  sort={Dens}
}

\newglossaryentry{densMat}%
{%
  name={\ensuremath{\widehat{\rho}}},
  description={The density matrix of the canonical ensemble, $\exp(-\gls{invTemp}\gls{sysHam})$},
  sort={Dens}
}

\newglossaryentry{hamDens}%
{%
  name={\ensuremath{\hat{h}}},
  description={The local hamiltonian},
  sort={Ham}
}

\newglossaryentry{sysHam}%
{%
  name={\ensuremath{\widehat{{\cal H}}}},
  description={The unperturbed system hamiltonian},
  sort={Ham}
}

\newglossaryentry{freeHam}%
{%
  name={\ensuremath{\widehat{H}_0}},
  description={The hamiltonian of the free phonon system},
  sort={Ham}
}

\newglossaryentry{intHam}%
{%
  name={\ensuremath{\widehat{H}_i}},
  description={The interaction hamiltonian},
  sort={Ham}
}

\newglossaryentry{pertHam}%
{%
  name={\ensuremath{\widehat{H}'}},
  description={The perturbing hamiltonian},
  sort={Ham}
}

\newglossaryentry{heatFluxOp}%
{%
  name={\ensuremath{\widehat{\mathbf{S}}}},
  description={The heat-flux operator, as derived by Hardy},
  sort={Ham}
}

\newglossaryentry{heatFluxOpd}%
{%
  name={\ensuremath{\widehat{S}}},
  description={A component of the heat-flux operator},
  sort={Ham}
}

\newglossaryentry{momOp}%
{%
  name={\ensuremath{\widehat{\mathbf{p}}}},
  description={The momentum operator of a particle in the crystal},
  sort={Mom}
}

\newglossaryentry{momOpd}%
{%
  name={\ensuremath{\widehat{p}}},
  description={A component of the momentum operator of a particle in the crystal},
  sort={Mom}
}

\newglossaryentry{phonMom}%
{%
  name={\ensuremath{\mathbf{q}}},
  description={The lattice momentum of a phonon in the crystal},
  sort={Mom}
}

\newglossaryentry{elMom}%
{%
  name={\ensuremath{\mathbf{k}}},
  description={The lattice momentum of an electron in the crystal},
  sort={Mom}
}

\newglossaryentry{ebi}%
{%
  name={m},
  description={The electron band index.},
  sort={Mom}
}

\newglossaryentry{posOp}%
{%
  name={\ensuremath{\widehat{\mathbf{r}}}},
  description={The position operator of a particle in the crystal},
  sort={Loc}
}

\newglossaryentry{posOpd}%
{%
  name={\ensuremath{\widehat{r}}},
  description={A component of the position operator of a particle in the crystal},
  sort={Loc}
}

\newglossaryentry{dispOp}%
{%
  name={\ensuremath{\widehat{\mathbf{u}}}},
  description={The displacement operator of a particle in the crystal},
  sort={Loc}
}

\newglossaryentry{dispOpd}%
{%
  name={\ensuremath{\widehat{u}}},
  description={A component of the displacement operator of a particle in the crystal},
  sort={Loc}
}

\newglossaryentry{enPol}%
{%
  name={\ensuremath{\widehat{\bm{{\cal P}}}}},
  description={The operator of \glqq{}energetic polarization\grqq{}},
  sort={enP}
}

\newglossaryentry{enPold}%
{%
  name={\ensuremath{\widehat{{\cal P}}}},
  description={A component of the operator of \glqq{}energetic polarization\grqq{}},
  sort={enP}
}

\newglossaryentry{potEn}%
{%
  name={\ensuremath{\widehat{V}}},
  description={The operator of potential energy. If it carries an index, it means an ions potential energy},
  sort={enP}
}

\newglossaryentry{locPotEn}%
{%
  name={\ensuremath{\widehat{\mathcal{V}}}},
  name={\ensuremath{\hat{v}}},
  description={The operator of local potential energy},
  sort={enP}
}

\newglossaryentry{locKinEn}%
{%
  name={\ensuremath{\widehat{\mathcal{V}}}},
  name={\ensuremath{\hat{t}}},
  description={The operator of local kinetic energy},
  sort={enP}
}

\newglossaryentry{genOp}%
{%
  name={\ensuremath{\widehat{\mathcal{O}}}},
  description={The placeholder for any operator. Usually this means a monomial of momentum and displacement operators of the phonons},
  sort={Loc}
}

\newglossaryentry{elCrea}%
{%
  name={\ensuremath{\widehat{c}^\dagger}},
  description={The electron creation operator},
  sort={Loc}
}

\newglossaryentry{elDest}%
{%
  name={\ensuremath{\widehat{c}}},
  description={The electron annihilation operator},
  sort={Loc}
}

\newglossaryentry{phonCrea}%
{%
  name={\ensuremath{\widehat{a}^\dagger}},
  description={The phonon creation operator},
  sort={Loc}
}

\newglossaryentry{phonDest}%
{%
  name={\ensuremath{\widehat{a}}},
  description={The phonon destruction operator},
  sort={Loc}
}

\newglossaryentry{phonMomOp}%
{%
  name={\ensuremath{\widehat{P}}},
  description={The momentum operator for a phonon in the crystal},
  sort={Loc}
}

\newglossaryentry{phonDispOp}%
{%
  name={\ensuremath{\widehat{U}}},
  description={The displacement operator for a phonon in the crystal},
  sort={Loc}
}

\newglossaryentry{elecProp}%
{%
  name={\ensuremath{\mathcal{D}}},
  description={The propagator of the electron},
  sort={Loc}
}

\newglossaryentry{phonProp}%
{%
  name={\ensuremath{\mathcal{G}}},
  description={The propagator of the phonon},
  sort={Loc}
}

\newglossaryentry{freePhonProp}%
{%
  name={\ensuremath{\mathcal{G}^{(0)}}},
  description={The propagator of the free phonon},
  sort={Loc}
}

\newglossaryentry{mixProp}%
{%
  name={\ensuremath{\mathcal{G}^{\gls{phonDispOp}\gls{phonMomOp}}}},
  description={The contraction of the phonon momentum and displacement operator},
  sort={Loc}
}

\newglossaryentry{freeMixProp}%
{%
  name={\ensuremath{\mathcal{G}^{\gls{phonDispOp}\gls{phonMomOp}(0)}}},
  description={The unperturbed contraction of the phonon momentum and displacement operator},
  sort={Loc}
}

\newglossaryentry{thec}%
{%
  name={\ensuremath{\kappa^{ij}}},
  description={The thermal conductivity of the system of interest},
  sort={enP}
}
\newglossaryentry{HFACF}%
{%
  name={\ensuremath{\Pi_\text{ret}^{ij}(E)}},
  description={The retarded heat-flux autocorrelation function of the system},
  sort={enP}
}

\newglossaryentry{HFACFt}%
{%
  name={\ensuremath{\Pi_\text{ret}^{ij}}},
  description={The real time retarded heat-flux autocorrelation function of the system},
  sort={enP}
}

\newglossaryentry{HFACFM}%
{%
  name={\ensuremath{\Pi_\text{M}^{ij}(E)}},
  description={Fourier form of Matsubara heat-flux autocorrelation function of the system},
  sort={enP}
}

\newglossaryentry{HFACFMt}%
{%
  name={\ensuremath{\Pi_\text{M}^{ij}}},
  description={The Matsubara heat-flux autocorrelation function of the system},
  sort={enP}
}

\newglossaryentry{matT}%
{%
  name={\ensuremath{T_\tau}},
  description={Matsubara time ordering operator},
  sort={Vol}
}

\newcommand{\eg}{{\it e.g.}, }
\newcommand{\ie}{{\it i.e.}, }

\begin{document}

\preprint{APS/123-QED}

\title{Gauge Invariance of the Thermal Conductivity in the Quantum Regime}

\author{Axel H\"ubner}
 \email{huebner@physik.hu-berlin.de}
\author{Santiago Rigamonti}
\author{Claudia Draxl}
\affiliation{%
 Humboldt-Universit\"at zu Berlin and IRIS Adlershof, Zum Gro{\ss}en Windkanal 2, D-12489 Berlin, Germany
}%

\date{\today}

\begin{abstract}
The widely used linear-response (LR) theory of thermal conduction in the quantum regime rests on the yet unproven assumption, that the thermal conductivity is invariant with respect to the gauge of the energy density of the system. This assumption manifests itself clearly in, \eg Hardy's formulation of the heat-flux operator [Phys. Rev. \textbf{132}, 168 (1963)]. In this work, we rigorously prove this assumption. Our proof, being valid for the nuclear and electronic subsystems, not only puts the state-of-the-art theory on solid grounds, but also enables going beyond the scope of the widely-used Boltzmann transport equation (BTE) within the LR framework for many-body systems. 
\end{abstract}

\maketitle

Linear-response (LR) theory is one of the most important techniques of theoretical physics to calculate materials properties \cite{kubo1957nonthermal, mahan2013many} such as optical  spectra \cite{wiser1963dielectric, adler1962quantum}, electron-phonon coupling strength \cite{giustino2017electron, baroni2001phonons}, magnetic susceptibility
\cite{lifshitz1956theory} and thermal conductivity \cite{kubo1957statistical, degroot1962non, hardy1963energy, nolting2008fundamentals}. The thermal conductivity of a material determines the amount of heat that flows through it in presence of a temperature gradient $\nabla T$. It is an important property for applications, such as thermographic defect detection \cite{almond2012analytical}, thermal-barrier coatings \cite{vassen2010overview}, and for thermoelectrics, that can tap waste heat to generate electricity \cite{snyder2011complex}. 

In crystals, heat is carried by electrons and lattice vibrations (phonons). To calculate the lattice contribution, the state-of-the-art methodologies are the Green-Kubo (GK) \cite{kubo1957statistical, qiu2009molecular} and the Boltzmann transport equation (BTE) \cite{Nag1980, Altland, broido2007intrinsic}. Despite recent advancements in the application of GK to real materials \cite{carbogno2017ab}, using it as a general framework for calculating heat transport is still a formidable task \cite{chen2010improve}. The BTE, therefore, remains the workhorse for computing the lattice thermal conductivity. Similarly, the electronic contribution to the thermal conductivity can be calculated using the Kubo-Greenwood approach \cite{ Holst2011} or the Boltzmann transport equation  \cite{Ziman2001electrons} where due to the computational complexity of the former, likewise the BTE remains the most often used technique.

The assumption underlying the BTE is that the heat-carrying particles, \ie phonons or electrons, act as in a gas, in which they do not scatter very often and behave almost like classical particles \cite{Altland}. This enables one to treat them (almost) independently and assign them a defined long lifetime which leads to the relaxation-time approximation (RTA) \cite{Ziman2001electrons}. Having lifetime and dispersion, one can calculate the respective contribution to the thermal conductivity of a material \cite{ThermPropGRIMVALL}. The BTE gives very good results for many materials \cite{chen2005monte, lacroix2006monte,zhou2014lattice, tadano2015self, carbogno2017ab,protik2020coupled}.  However, this method becomes questionable if the scattering rates of the particles are large \cite{pines2018theory}, as, for instance, phonons in materials with low lattice thermal conductivity. 

The BTE can be derived as an approximation to Kubo's formula \cite{kubo1957statistical, luttinger1964theory} (Eq. \eqref{eq:kappadef} below) for electronic as well as lattice thermal transport \cite{schieve1962correlation, mahan2013many}. Therefore, it should be possible to derive correction terms that may answer the question \textit{when} the BTE is applicable to a material, and if not, \textit{how} to improve it. However, this approach faces a severe obstacle: For thermal transport, a temperature gradient is required. Therefore, to formulate the LR, small subsystems in local equilibrium are considered \cite{allen1993thermal}, which gives rise to the notion of a local energy density \cite{mahan2013many}. For interacting systems, this local energy density is, however, ambiguous: Whenever two or more particles interact, portions of the interaction energy can be arbitrarily assigned to any of them, or even to other points in space \cite{hardy1963energy}. As a consequence, its definition involves the choice of a gauge. This is the case, for instance, for the interacting electrons and atomic nuclei responsible for the 
thermal conduction. 

Even though it has been argued long ago that the LR approach is applicable also in the presence of a thermal disturbance ($\nabla T \neq 0$) \cite{kubo1957statistical,luttinger1964theory}, it remained unclear, whether the result is invariant with respect to the specific definition, \ie the gauge, of the local energy density. Only for the classical case a proof has been given a few years ago \cite{marcolongo2016microscopic}. In contrast, however, the many-body system of interacting particles that form any material, is a quantum problem. This is where our work comes in.

In this Letter, we prove the gauge invariance of the thermal conductivity in the quantum regime. In analogy to the classical counterpart \cite{marcolongo2016microscopic}, here we derive a quantum Einstein relation, which relates the thermal conductivity to the equal-time autocorrelation of a quantum energy-displacement operator. This enables us to leverage a generalized version of the Cauchy-Schwarz inequality to show that, for any two gauges, the thermal conductivity tensors must be identical. We then use Hardy's formulation of the heat flux operator \cite{hardy1963energy} to  analyze the thermodynamic limit and illustrate the impact of the gauge. Finally, we show how the gauge invariance is fulfilled in a simple case of two-body interactions. This example motivates an alternative proof that is given at the end.

The lattice thermal conductivity is given as \cite{mahan2013many, kubo1957statistical, allen1993thermal}

\begin{align} \label{eq:kappadef}
\gls{thec}(\omega) = \frac{\gls{sysVol}}{T} 
\int_0^\beta \!\! d\lambda
\int_0^\infty \!\!\!\!  dt e^{i (\omega+i\eta) t} & ~ \langle \gls{heatFluxOpd}^i  \gls{heatFluxOpd}^j (t + i\hbar\lambda) \rangle
\end{align}
where \gls{sysVol} is the system volume, \textit{T} the temperature, $\beta = (k_B T)^{-1}$, and $k_B$ the Boltzmann constant. $\eta$ is a positive infinitesimal. The operators are written in the Heisenberg picture. Angle brackets denote canonical thermal averages: $\langle \widehat{\mathcal{O}} \rangle = (\gls{tr} \gls{densMat})^{-1} \gls{tr} (\gls{densMat}\widehat{\mathcal{O}})$ with  $\gls{densMat} = \exp(-\beta \gls{sysHam})$ the density matrix, and \gls{sysHam} the system hamiltonian in equilibrium. $\gls{heatFluxOpd}^i$ are the cartesian components of the heat flux operator \gls{heatFluxOp}.  It can be obtained from the volume average of a {\it local} heat-flux operator $\widehat{\mathbf{s}}(\bm{x})$ which fulfills the continuity equation \cite{hardy1963energy}
\begin{align} \label{eq:cont}
\frac{1}{i\hbar} \left[\gls{hamDens}(\bm{x}),\gls{sysHam}\right] = - \nabla \cdot \widehat{\mathbf{s}}(\bm{x}),
\end{align}
where $\gls{hamDens}(\bm{x})$ is a hamiltonian density that is only constrained by the fact that its volume integral is \gls{sysHam}. There remains a gauge freedom for \gls{hamDens}($\bm{x}$), since the interaction energy can be located at any of the involved particles or spatial coordinates. Multiplying Eq. \eqref{eq:cont} with $\bm{x}$ and integrating over space, using partial integration, leads to
\begin{align} \label{eq:partialInt}
\frac{1}{i\hbar} \left[\gls{enPol},\gls{sysHam}\right] = - \frac{1}{\gls{sysVol}}\int_{\gls{sysVol}} d\bm{x}~ \bm{x} \nabla \cdot \widehat{\mathbf{s}}(\bm{x}) = \gls{heatFluxOp}.
\end{align}
Here,
\begin{align} \label{eq:epol}
    \gls{enPol}=\frac{1}{\gls{sysVol}}\int_{\gls{sysVol}} d\bm{x}~ \bm{x} \gls{hamDens}(\bm{x})
\end{align}
is the (gauge-dependent) energy polarization per unit volume. The surface term in Eq. \eqref{eq:partialInt} vanishes since the integrand $\bm{x} \widehat{\mathbf{s}}(\bm{x})$ is zero outside the system \cite{hardy1963energy}. From Eq. \eqref{eq:epol}, it is clear that different choices of the gauge in $\gls{hamDens}(\bm{x})$ lead in general to different definitions of \gls{heatFluxOp}. While it is clear, that a physical observable is indeed independent of arbitrary mathematical gauge choices, the question remains whether the result of LR theory is also independent. In what follows, we rigorously prove that the LR result for \gls{thec} is indeed independent of the gauge of $\gls{hamDens}(\bm{x})$.  

We define the quantum energy displacement
\begin{align} \label{eq:aux}
\widehat{D}^i (t) =  \left(\frac{\Omega k_B}{2}\right)^{1/2} \int_{-\gls{invTemp}/2}^{\gls{invTemp}/2} \!\! d\lambda \int_0^t \!\! dt'  \gls{heatFluxOpd}^i(t'-i\hbar\lambda).
\end{align}
Now, we relate the thermal conductivity \gls{thec} to $\text{Re}\langle \widehat{D}^i (t) \widehat{D}^j (t)\rangle$. With the aid of \cite{helfand1960transport} 
\begin{align}\label{eqn:helfand}
\int_{c}^{c+y} \!\! dx_1 dx_2 ~ f(x_1-x_2) = \int_{-y}^y \!\! d\gamma~ (y-|\gamma|) f(\gamma) ,
\end{align}
where $c$ is some constant, we find
\begin{align} \label{eq:DD}
\langle & \widehat{D}^i (t) \widehat{D}^j (t) \rangle = \nonumber \\
&  \frac{\Omega k_B}{2} \!\!\!\int_{-\gls{invTemp}}^{\gls{invTemp}} \!\!\!\! d\phi (\beta - |\phi|) \int_{-t}^t \!\!\!\!d\sigma (t-|\sigma|) \langle \gls{heatFluxOpd}^i(\sigma-i\hbar\phi)
\gls{heatFluxOpd}^j \rangle.
\end{align}
Applying the relation
\begin{align}
\lim_{t\rightarrow\infty} \int_{-t}^t d\sigma \left(1-\frac{|\sigma|}{t}\right) \exp(i\frac{\mathcal{E} \sigma}{\hbar}) = 2\pi \hbar \delta (\mathcal{E})
\end{align}
to Eq.\eqref{eq:DD} in the eigenfunction representation of \gls{sysHam} and taking the limit $t \rightarrow \infty$, we arrive at
\begin{align} \label{eq:eiglimit}
 \lim_{t \rightarrow \infty} \frac{1}{t} &  \langle \widehat{D}^i (t) \widehat{D}^j (t) \rangle  =  \nonumber \\ 
& \frac{ \pi \hbar \Omega k_B \gls{invTemp}^2 }{\gls{tr}(\gls{densMat})} \sum_{\gamma,\xi} e^{-\gls{invTemp} \gls{eigene}_\gamma} \delta(\gls{eigene}_{\gamma\xi})\gls{heatFluxOpd}^i_{\gamma,\xi}\gls{heatFluxOpd}^j_{\xi,\gamma} \, ,
\end{align}
where $\gls{heatFluxOpd}^i_{\gamma,\xi}$ is a matrix element of the heat-flux operator between \gls{sysHam}'s eigenstates $|\xi \rangle$ and $|\gamma\rangle$, $\gls{eigene}_{\xi/\gamma}$ are the corresponding eigenvalues, and $\gls{eigene}_{\gamma\xi}=\gls{eigene}_{\gamma}-\gls{eigene}_\xi$. This limit is identical to the static thermal conductivity \gls{thec}, as obtained from the real part of Eq.\eqref{eq:kappadef} in the $\omega \rightarrow 0$, then $\eta \rightarrow 0$ limit. Therefore
\begin{align} 
\lim_{t\rightarrow\infty} \frac{1}{t} \gls{re} \langle \widehat{D}^i (t) \widehat{D}^j (t) \rangle = \kappa^{ij}.\label{eq:ddkappa}
\end{align}
This is the quantum Einstein relation that we wanted to prove.

Now, only one more step is required: Assume there are two different heat fluxes, $\widehat{S}_1^{i}$ and $\widehat{S}_2^{i}$ (corresponding to two different gauges), causing two different conductivities, $\kappa_1^{ii}$ and $\kappa_2^{ii}$, as obtained from Eq. \eqref{eq:ddkappa} with $\widehat{D}^i_1$ and $\widehat{D}^i_2$, respectively. We define $\widehat{\Psi}^i=\widehat{D}^i_2 - \widehat{D}^i_1 $. Then
\begin{align}\label{eqn:combTc}
\kappa_{12}^{ii} &= \lim_{t\rightarrow\infty} \frac{1}{t} \gls{re} \langle (\widehat{D}^i_1(t)+\widehat{D}^i_2(t))^2 \rangle \nonumber \\
&= \kappa_1^{ii} + \kappa_2^{ii} + \lim_{t\rightarrow\infty} \frac{1}{t} \gls{re} \langle \widehat{D}^i_1(t)\widehat{D}^i_2(t) +\widehat{D}^i_2(t)\widehat{D}^i_1(t) \rangle \nonumber \\
&= 2~ (\kappa_1^{ii} + \kappa_2^{ii}) - \lim_{t\rightarrow\infty} \frac{1}{t} \gls{re} \langle \widehat{\Psi}^i(t) \widehat{\Psi}^i(t) \rangle \nonumber \\
&= 2~ (\kappa_1^{ii} + \kappa_2^{ii}).
\end{align}
The last step is possible, because
\begin{align}
    \lim_{t\rightarrow\infty} \frac{1}{t} \langle \widehat{\Psi}^i(t) \widehat{\Psi}^i(t) \rangle = 0. \label{eq:psipsi}
\end{align}
We will elaborate on this point further below. Comparing the second with the fourth line, we find
\begin{align}
\lim_{t\rightarrow\infty} \frac{1}{t} \langle \widehat{D}^i_1\widehat{D}^i_2 +\widehat{D}^i_2\widehat{D}^i_1 \rangle =  \kappa_1^{ii} + \kappa_2^{ii}.
\end{align}
Using the Cauchy-Schwarz inequality for hermitian operators $\widehat{A}$ and $\widehat{B}$ 
\begin{align}\label{eqn:startCSI}
4 ~ \langle \widehat{A}^2 \rangle \langle \widehat{B}^2 \rangle \geq \langle \widehat{A}\widehat{B}+\widehat{B}\widehat{A} \rangle^2 \; ,
\end{align}

we arrive at:
\begin{align} \label{eqn:kappaInv}
2 \sqrt{\kappa_1^{ii}\kappa_2^{ii}} \geq  \kappa_1^{ii} + \kappa_2^{ii} \nonumber\\
\Rightarrow 0 \geq (\kappa_1^{ii} - \kappa_2^{ii})^2 \; .
\end{align}
The invariance of the tensor \gls{thec} follows immediately from Eq. \eqref{eqn:kappaInv} which is true for all directions, not just the cartesian axes. Since the tensors $\mathbf{\kappa}_1$ and $\mathbf{\kappa}_2$ are real and symmetric  \cite{de1951thermodynamics}, their difference $\theta = \kappa_1-\kappa_2$ is also symmetric and has an eigenbasis. The representation of the tensor $\theta$ in the coordinate system defined by this eigenbasis must be exactly zero, according to Eq. \eqref{eqn:kappaInv}. As it vanishes in this basis, it vanishes in general. Therefore, $\mathbf{\kappa}_1 = \mathbf{\kappa}_2$. $\blacksquare$

An important step in the above derivation is Eq.(\ref{eq:psipsi}). We will now prove it. Replacing $\widehat{D}^i$ with $\widehat{\Psi}^i$ in Eq. \eqref{eq:DD}, one finds
\begin{align}
\lim_{t\rightarrow\infty}  \frac{1}{t} \langle \widehat{\Psi}^i (t) &\widehat{\Psi}^i (t) \rangle = 
\nonumber \\ 
& \hspace{-0.05cm} \frac{ \pi  \Omega k_B \gls{invTemp}^2 }{\hbar \gls{tr}(\gls{densMat})} \sum_{\gamma,\xi} e^{-\gls{invTemp} \gls{eigene}_\gamma} \delta(\gls{eigene}_{\gamma\xi})|\Delta\gls{enPold}_{\gamma,\xi}^i|^2  \gls{eigene}_{\gamma\xi}^2 \; .
\label{eq:psipsieig}
\end{align}
Here, $\Delta\gls{enPold}_{\gamma,\xi}^i$ are matrix elements of the $i$ component of $\Delta\gls{enPol} = \gls{enPol}_2-\gls{enPol}_1$, \textit{i.e.} the change in energy polarization $\gls{enPol}$ due to the change of gauge. This quantity must be bounded for physical gauges. A way to understand this is to consider the relocation of a part of the energy $\delta \widehat{v}(\bm{x})$ of a subsystem $A$ of volume $\Omega_A$ by a constant vector $\bm{c}$, such that the hamiltonian density is changed to $\widehat{h}'(\bm{x}) = \widehat{h}(\bm{x}) + \Delta \widehat{h}(\bm{x})$,
\begin{align}
    \Delta \widehat{h}(\bm{x}) =  \vartheta_A(\bm{x}-\bm{c})  \;\! \delta \;\! \widehat{v}(\bm{x}-\bm{c}) - \vartheta_A(\bm{x}) \;\! \delta \;\! \widehat{v}(\bm{x}) ,
\end{align}
where $\vartheta_A(x)$ is an indicator function that is one in the volume $\Omega_A$ and otherwise zero. Then, the change of polarization is given by
\begin{align} \label{eq:epolchange}
    \Delta \gls{enPol} = \int d\bm{x} \;\!  \bm{x} \;\!  \Delta\widehat{h}(\bm{x}) = \frac{\bm{c}}{\gls{sysVol}} \int_{\Omega_A} d\bm{x} ~ \delta \;\!  \widehat{v}(\bm{x}).
\end{align}
Under the premises, that the relocation $\bm{c}$ of the energy should not be macroscopically large \cite{allen1993thermal} and that the total energy of a subsystem should remain finite, $\Delta \gls{enPol}$ must be bounded. As $\Delta \gls{enPol}$ is bounded, Eq. \eqref{eq:psipsieig} equals to zero. This confirms the validity of Eq. \eqref{eq:psipsi} and the proof is complete.

Before passing to an example, we discuss a few aspects concerning the thermodynamic limit and the heat flux operator. In the thermodynamic limit ($\Omega\rightarrow\infty$), the energy polarization as given by Eq. \eqref{eq:epol} is ill-defined, due to the presence of $\bm{x}$ in the integral. Thus, in this limit, no meaningful expression for \gls{thec} in terms of \gls{enPol} can be written,  as would, \eg be obtained by replacing \gls{heatFluxOp} in Eq. \eqref{eq:eiglimit} by Eq. \eqref{eq:partialInt}:
\begin{equation} \gls{thec} = \hspace{-0.05cm} \frac{ \pi  \Omega k_B \gls{invTemp}^2 }{\hbar \gls{tr}(\gls{densMat})} \sum_{\gamma,\xi} e^{-\gls{invTemp} \gls{eigene}_\gamma} \delta(\gls{eigene}_{\gamma\xi})\gls{eigene}_{\gamma\xi}^2 \gls{re}(\gls{enPold}_{\gamma,\xi}^i \gls{enPold}_{\xi,\gamma}^j)   \; .\label{eq:kappap} 
\end{equation}

One may argue that a realistic system will never be of infinite size, and employ the above equation. In such a case, though, the term $\delta(\gls{eigene}_{\gamma\xi})\gls{eigene}_{\gamma\xi}^2$ would seemingly lead to zero conductivity. However, for a finite system the interaction with the environment must be taken into consideration. This interaction, which enables the transfer of energy through the material, leads to a broadening of $\delta(\gls{eigene}_{\gamma\xi})$ and thereby to a finite thermal conductivity. 

In contrast, the analogous expression in terms of the heat flux operator, namely
\begin{align}\label{eq:kappawiths}
\gls{thec} &= \hspace{-0.05cm} \frac{ \pi \hbar \Omega k_B \gls{invTemp}^2 }{ \gls{tr}(\gls{densMat})} \sum_{\gamma,\xi} e^{-\gls{invTemp} \gls{eigene}_\gamma} \delta(\gls{eigene}_{\gamma\xi})\gls{re}(\gls{heatFluxOpd}_{\gamma,\xi}^i \gls{heatFluxOpd}_{\xi,\gamma}^j)\;,
\end{align}
is well defined also in the thermodynamic limit. This can be understood from the general expression for $\gls{heatFluxOp}$ as was found by Hardy \cite{hardy1963energy}
\begin{align}\label{eqn:defHeatFluxOp} 
\gls{heatFluxOp} =  \frac{1}{2\gls{sysVol}}  \sum_{\gls{part1}} \frac{\gls{momOp}_{\gls{part1}}}{M_{\gls{part1}}}\widehat{H}_l - \frac{i}{2\gls{sysVol}\hbar} \sum_{\gls{part1},\gls{part2}} \gls{posOp}_{\gls{part1}\gls{part2}} \left[\widehat{T}_{\gls{part1}},\gls{potEn}_{\gls{part2}}\right] +\text{h.c.}
\end{align}
Here, $\widehat{H}_{\gls{part1}}=\widehat{T}_{\gls{part1}}+\gls{potEn}_{\gls{part1}}$, $\widehat{T}_l=\gls{momOp}_{\gls{part1}}^2/2M_{\gls{part1}}$, $\gls{posOp}_{\gls{part1}\gls{part2}} = \gls{posOp}_{\gls{part1}}-\gls{posOp}_{\gls{part2}}$,  $\gls{momOp}_{\gls{part1}}$ and $\gls{posOp}_{\gls{part1}}$ are the momentum and position operators of particle \gls{part1}, $M_{\gls{part1}}$ is its mass, and  $\gls{potEn}_{\gls{part1}}$ its potential energy. As will be shown in the example below, the definition of the latter depends on the choice of the gauge, the only constraint being that the sum of the $\gls{potEn}_{\gls{part1}}$ must equal the potential-energy operator of the system, \gls{potEn}. The first term of Eq. \eqref{eqn:defHeatFluxOp} can be interpreted as the transport of heat carried by the particle flux, and is dominant if the particles form a gas, as in the case of electrons \cite{allen1993thermal, Holst2011, hardy1963energy}. The second term can be interpreted as the rate of work that particle \gls{part2} does on particle \gls{part1} \cite{allen1993thermal},  and dominates the lattice transport. Similar (approximate) operators for the thermal current, can be found in literature, \eg \cite{luttinger1964theory, Zlatic2014}. 

The main reason why \gls{heatFluxOp}, as given by Eq. \eqref{eqn:defHeatFluxOp}, can be employed in Eq. \eqref{eq:kappawiths} in the thermodynamic limit, relies on the fact that it only contains position differences, $\gls{posOp}_{\gls{part1}\gls{part2}}$. This is in contrast to \gls{enPol}, which contains absolute positions, $\bm{x}$. Moreover, analogous premises as given after Eq. \eqref{eq:epolchange} apply here: (i) No gauge choice should lead to divergent local energies $\widehat{H}_{\gls{part1}}$, and (ii) interactions should not have a macroscopic range, \ie the energy cannot be relocated over macroscopic distances. More precisely, as the particle separation $d$ increases, the term $\gls{posOp}_{\gls{part1}\gls{part2}} \left[\widehat{T}_{\gls{part1}},\gls{potEn}_{\gls{part2}}\right]$ must vanish faster than $1/d^3$.  This is the case, for instance, for the nuclei in non-polar insulators and most metals, for which the interatomic force constants decay as $d^{-5}$ or faster \cite{gonze1997dynamical}. For polar insulators and semiconductors, the interatomic force constants decay like $d^{-3}$ but due to the total charge neutrality of the system, they oscillate, leading to a bounded heat-flux operator. This is more clear in reciprocal space \cite{hardy1963energy} as the heat-flux operator is well-defined also in presence of LO-TO splitting \cite{gonze1997dynamical}.

We exemplify our proof with a system that has only two-body interactions between the particles. An analogous example was considered in Ref. \cite{Baroni2020} for the classical case. For this system, the local energy density $\gls{hamDens}(\bm{x})$ can be written as 
\begin{align}
    \gls{hamDens}(\bm{x}) = \frac{1}{2} \sum_{\gls{part1}} \left[ \delta(\bm{x} - \gls{posOp}_{\gls{part1}}) \widehat{H}_{\gls{part1}}(\Gamma) + \widehat{H}_{\gls{part1}}(\Gamma) \delta(\bm{x} - \gls{posOp}_{\gls{part1}}) \right] 
\end{align}
where $\widehat{H}_{\gls{part1}}(\Gamma)$ is the local energy of particle \gls{part1}, given as $\widehat{H}_{\gls{part1}}(\Gamma)=\widehat{T}_{\gls{part1}}+\widehat{V}_{\gls{part1}}(\Gamma)$, with
\begin{align}\label{eq:vijexplicit}
    \widehat{V}_{\gls{part1}}(\Gamma) = \frac{1}{2} \sum_{\gls{part2} \neq \gls{part1}}  \widehat{v}_{\gls{part1}\gls{part2}} (1 + \Gamma_{\gls{part1}\gls{part2}}).
\end{align}
Here, $\widehat{v}_{\gls{part1}\gls{part2}}$ is the interaction operator between particles \gls{part1} and \gls{part2} that depends on their positions. The assumptions after Eq. \eqref{eqn:defHeatFluxOp} regarding the interaction range apply to $\widehat{v}_{\gls{part1}\gls{part2}}$ as well. $\Gamma$ is an arbitrary antisymmetric matrix, thus $\sum_{\gls{part1}} \widehat{V}_{\gls{part1}}(\Gamma)=\widehat{V}$, independently of $\Gamma$, as $ \widehat{v}_{\gls{part1}\gls{part2}} =  \widehat{v}_{\gls{part2}\gls{part1}}$. In other words, $\Gamma$ parametrizes the gauge choice. For instance, if $\Gamma_{\gls{part1}\gls{part2}}=\text{sgn}(\gls{part1}-\gls{part2})$, all the interaction energy between particles \gls{part1} and \gls{part2} is assigned to the particle with index $\min(\gls{part1},\gls{part2})$. By plugging Eq. \eqref{eq:vijexplicit} into Eq. \eqref{eqn:defHeatFluxOp}, we obtain the change in the heat-flux operator due to the application of a gauge $\Gamma$, \ie $\Delta \gls{heatFluxOp} (\Gamma)=\gls{heatFluxOp} (\Gamma)-\gls{heatFluxOp} (\Gamma=0)$. Then, using $\Delta \gls{heatFluxOp} (\Gamma)= d\Delta \gls{enPol} (\Gamma)/dt$ (cf. Eq.(\ref{eq:partialInt})), we get the corresponding change in energy polarization (up to a time-independent constant),
\begin{align}    
\Delta \gls{enPol} (\Gamma) & = \frac{1}{2 \gls{sysVol}}  \sum_{\gls{part1}, \gls{part2}>\gls{part1}} \gls{posOp}_{\gls{part1}\gls{part2}} \widehat{v}_{{\gls{part1}}{\gls{part2}}} \Gamma_{\gls{part1}\gls{part2}} + \text{h.c.} .\nonumber
\end{align}
Here, the spatial relocation of the energy is limited by the range of the interaction $\widehat{v}_{\gls{part1}{\gls{part2}}}$, which plays a role analogous to the vector $\bm{c}$ in Eq. \eqref{eq:epolchange}. This result completes the example, since the energy polarization difference $\Delta \gls{enPol} (\Gamma)$ is bounded, and, therefore, Eq. \eqref{eq:psipsi} is fulfilled (see Eq. \eqref{eq:psipsieig}).

From this example, we note that, interestingly, the difference in \gls{thec} for two different definitions of \gls{heatFluxOp} contains at least one  $\Delta \gls{heatFluxOp}$ term, as can be deduced from Eq.(\ref{eq:kappawiths}). Therefore, by using $\Delta \gls{heatFluxOp} = \frac{1}{i\hbar} \left[\Delta \gls{enPol},\gls{sysHam}\right]$, the gauge related change in \gls{thec} will be
\begin{align}\label{eqn:bb}
 - \frac{ i \pi \Omega k_B \gls{invTemp}^2 }{\gls{tr}(\gls{densMat})} \sum_{\gamma,\xi} \gls{re}(\Delta\gls{enPold}^i_{\gamma\xi} \gls{heatFluxOpd}^j_{\xi\gamma}) e^{-\gls{invTemp} \gls{eigene}_\gamma} \delta(\gls{eigene}_{\gamma\xi}) \gls{eigene}_{\gamma\xi} = 0.
\end{align}
which is zero because $x \delta(x) = 0$ and both $\Delta \gls{enPol}$ and $\gls{heatFluxOp}$ are well defined in the thermodynamic limit. This may be considered as a condensed version of our proof above.

To conclude, we have shown that the thermal conductivity is invariant of the specific choice of gauge in the energy density. This guarantees that one obtains its unique, true value in the realm of linear-response theory. With this, derivations of correction terms to the Boltzmann transport equation for thermal conductivity become possible, as any gauge of the energy density can be used. The correction terms, as proposed for example in Ref. \cite{sham1967temperature}, will enable one to check for which materials the Boltzmann transport equation can be used and for which it needs to be improved. This has been a long-standing, important issue for applied materials, such as thermoelectrics. Last but not least, we emphasize that, following our procedure, the gauge invariance for other quantities related to the heat-flux like the Seebeck coefficient, may be derived.

\section{Acknowledgements}
Work supported by the Studienstiftung des Deutschen Volkes. The authors thank Keith Gilmore, Nakib H. Protik and Pasquale Pavone for fruitful discussions.

\end{document}